\documentclass[aps,reprint,superscriptaddress,nofootinbib]{revtex4-2}

\usepackage{graphicx}% Include figure files
\usepackage{dcolumn}% Align table columns on decimal point
\usepackage{bm}% bold math
\usepackage[colorlinks=true,  urlcolor=blue,  linkcolor=blue,  citecolor=blue]{hyperref}% add hypertext capabilities

\usepackage{amsmath, mathtools}
\usepackage{amssymb} % for \lessapprox notably
\usepackage{amsfonts,bbm}
\usepackage{stmaryrd}
\usepackage{upgreek}
\usepackage{comment}
\usepackage{multirow}
\usepackage{caption}
\usepackage{subcaption}
\usepackage[export]{adjustbox}

\renewcommand*\vec[1]{\bm{#1}}
\newcommand{\ve}[1]{{\vec{#1}}}

\newcommand\dI{\mathrm{d}}

\newcommand*{\conjb}[1]{\overline{#1}}

\providecommand{\sorthelp}[1]{} % For Planck_bib.bib

\let\jnlstyle=\rmfamily 
\def\refjnl#1{{\jnlstyle#1}}% 
\newcommand\pnas{\refjnl{PNAS}}
\newcommand\jmlr{\refjnl{JMLR}}
\newcommand\aj{\refjnl{AJ}}%        % Astronomical Journal 
\renewcommand\apj{\refjnl{ApJ}}%    % Astrophysical Journal 
\newcommand\apjs{\refjnl{ApJS}}%    % Astrophysical Journal, Supplement 
\newcommand\aap{\refjnl{A\&A}}%     % Astronomy and Astrophysics 
\newcommand\mnras{\refjnl{MNRAS}}%   % Monthly Notices of the RAS 
\renewcommand\prd{\refjnl{PhRvD}}% % Physical Review D 
\newcommand\pasj{\refjnl{PASJ}}%     % Publications of the ASJ 
\renewcommand\nat{\refjnl{Nature}}%  % Nature 
\newcommand\physrep{\refjnl{PhR}}%       % Physics Reports 
\newcommand\jcap{\refjnl{JCAP}}%  % Journal of Cosmology and Astroparticle Physics
\graphicspath{{figures/}}

\begin{document}

\title{Galaxy Clustering Analysis with {\sc SimBIG} and the Wavelet Scattering Transform}
%\thanks{A footnote to the article title}

%\author[0000-0003-0055-0953]{Bruno Régaldo-Saint Blancard}
\author{Bruno Régaldo-Saint Blancard}
\email{bregaldo@flatironinstitute.org}
\affiliation{Center for Computational Mathematics, Flatiron Institute, 162 5\textsuperscript{th} Avenue, New York, NY 10010, USA}

%\author[0000-0003-1197-0902]{ChangHoon Hahn}
\author{ChangHoon Hahn}
\affiliation{Department of Astrophysical Sciences, Princeton University, Princeton NJ 08544, USA} 

\author{Shirley Ho}
\affiliation{Center for Computational Astrophysics, Flatiron Institute, 162 5\textsuperscript{th} Avenue, New York, NY 10010, USA}

\author{Jiamin Hou}
\affiliation{Department of Astronomy, University of Florida, 211 Bryant Space Science Center, Gainesville, FL 32611, USA}
\affiliation{Max-Planck-Institut f\"ur Extraterrestrische Physik, Postfach 1312, Giessenbachstrasse 1, 85748 Garching bei M\"unchen, Germany}

\author{Pablo Lemos}
\affiliation{Department of Physics, Universit\'{e} de Montr\'{e}al, Montr\'{e}al, 1375 Avenue Th\'{e}r\`{e}se-Lavoie-Roux, QC H2V 0B3, Canada}
\affiliation{Mila - Quebec Artificial Intelligence Institute, Montr\'{e}al, 6666 Rue Saint-Urbain, QC H2S 3H1, Canada}

%\author[0000-0002-0637-8042]{Elena Massara}
\author{Elena Massara}
\affiliation{Waterloo Centre for Astrophysics, University of Waterloo, 200 University Ave W, Waterloo, ON N2L 3G1, Canada}
\affiliation{Department of Physics and Astronomy, University of Waterloo, 200 University Ave W, Waterloo, ON N2L 3G1, Canada}

\author{Chirag Modi}
\affiliation{Center for Computational Mathematics, Flatiron Institute, 162 5\textsuperscript{th} Avenue, New York, NY 10010, USA}
\affiliation{Center for Computational Astrophysics, Flatiron Institute, 162 5\textsuperscript{th} Avenue, New York, NY 10010, USA}

%\author[0000-0001-8841-9989]{Azadeh Moradinezhad Dizgah}
\author{Azadeh Moradinezhad Dizgah}
\affiliation{D\'epartement de Physique Th\'eorique, Universit\'e de Gen\`eve, 24 quai Ernest Ansermet, 1211 Gen\`eve 4, Switzerland}

\author{Liam Parker}
\affiliation{Department of Physics, Princeton University, Princeton, NJ 08544, USA}

%\author[0000-0002-0985-7233]{Yuling Yao}
\author{Yuling Yao}
\affiliation{Center for Computational Mathematics, Flatiron Institute, 162 5\textsuperscript{th} Avenue, New York, NY 10010, USA}

\author{Michael Eickenberg}
\affiliation{Center for Computational Mathematics, Flatiron Institute, 162 5\textsuperscript{th} Avenue, New York, NY 10010, USA}

\collaboration{{\sc SimBIG} Collaboration}

\date{\today}

\begin{abstract}

The non-Gaussisan spatial distribution of galaxies traces the large-scale structure of the Universe and therefore constitutes a prime observable to constrain cosmological parameters. We conduct Bayesian inference of the $\Lambda$CDM parameters $\Omega_m$, $\Omega_b$, $h$, $n_s$, and $\sigma_8$ from the BOSS CMASS galaxy sample by combining the wavelet scattering transform (WST) with a simulation-based inference approach enabled by the {\sc SimBIG} forward model. We design a set of reduced WST statistics that leverage symmetries of redshift-space data. Posterior distributions are estimated with a conditional normalizing flow trained on 20,000 simulated {\sc SimBIG} galaxy catalogs with survey realism. We assess the accuracy of the posterior estimates using simulation-based calibration and quantify generalization and robustness to the change of forward model using a suite of 2,000 test simulations. When probing scales down to $k_{\rm max}=0.5~h/\text{Mpc}$, we are able to derive accurate posterior estimates that are robust to the change of forward model for all parameters, except~$\sigma_8$. We mitigate the robustness issues with $\sigma_8$ by removing the WST coefficients that probe scales smaller than $k \sim 0.3~h/\text{Mpc}$. Applied to the BOSS CMASS sample, our WST analysis yields seemingly improved constraints obtained from a standard PT-based power spectrum analysis with $k_{\rm max}=0.25~h/\text{Mpc}$ for all parameters except $h$.
However, we still raise concerns on these results. The observational predictions significantly vary across different normalizing flow architectures, which we interpret as a form of model misspecification. This highlights a key challenge for forward modeling approaches when using summary statistics that are sensitive to detailed model-specific or observational imprints on galaxy clustering.

\end{abstract}

\maketitle

% Body
\section{Introduction}
\label{sec:intro}

The evolution of the Universe is remarkably well understood under the lens of the standard model of cosmology, or $\Lambda$CDM model. The spatial distribution of galaxies traces the large-scale structure of the Universe (LSS) and constitutes a prime observable to constrain the parameters of this model. In the past decades, spectroscopic galaxy surveys such as the Sloan Digital Sky Survey III: Baryon Oscillation Spectroscopic Survey~\citep[BOSS,][]{Eisenstein2011, Dawson2013} provided us with redshift measurements for over a million galaxies spanning an effective volume of several Gpc$^3$. Upcoming instruments, including the Dark Energy Spectroscopy Instrument~\citep{Aghamousa2016}, the Subaru Prime Focus Spectrograph~\citep{Takada2014}, the {\em Euclid} satellite~\citep{Laureijs2011}, and the Nancy Grace Roman Space Telescope~\citep{Spergel2015} will surpass these numbers by more than an order of magnitude. This tremendous amount of data will provide a unique opportunity to precisely measure the expansion and growth histories of the Universe.

Cosmological inference from the observed galaxy distribution is commonly formulated within a Bayesian framework. Given a prior distribution on the target parameters, a set of \textit{summary statistics} describing the galaxy clustering, and a model for the corresponding likelihood function, Bayesian inference yields a posterior distribution on the parameters. Typically, the likelihood is assumed to follow a Gaussian functional form, although this assumption is hardly justifiable~\citep{Hahn2019, Park2023}.\footnote{The functional form of the likelihood a priori depends directly on the summary statistics.} In terms of summary statistics, galaxy clustering analyses thus far have focused primarily on the galaxy power spectrum, $P_\ell(k)$~\citep[\emph{e.g.}][]{Kaiser1987, Hamilton1997, Peacock2001, Tegmark2006, Guzzo2008, Beutler2017, Ivanov2020, Chen2020}, using perturbation theory (PT)~\citep{Bernardeau2002, Desjacques2018} to model its dependency on the cosmological parameters. These standard analyses are limited to (large) scales in the linear and weakly nonlinear regime, where PT is valid. As a result, they do not exploit cosmological information available in the strongly nonlinear regime or in the non-Gaussian properties of the galaxy distribution.

The literature has investigated several summary statistics beyond $P_\ell$, including the bispectrum~\citep{Hahn2021, Damico2022, Philcox2022}, marked power spectrum~\citep{Massara2022}, and skew spectra~\citep{Hou2022}. These works firmly demonstrate that non-Gaussian information significantly improves the precision on cosmological constraints. Unfortunately, traditional PT-based analyses hardly extend to these statistics, which makes analytic connections to the cosmological parameters unclear if not intractable, especially in the nonlinear regime. On the other hand, simulation-based inference\footnote{SBI is also often called \textit{likelihood-free inference} or \textit{implicit likelihood inference} in the literature.} (SBI), which has grown in popularity in astronomy and cosmology, offers an alternative approach to model the likelihood function by leveraging high-fidelity numerical simulations, and thus avoiding strong assumptions on the likelihood. Typically, neural density estimators are trained on these simulations to enable efficient posterior sampling from the target observation~\citep{Alsing2019, Cranmer2020}.

In this context, \cite{Hahn2022a, Hahn2022b} introduced the {\sc SimBIG} forward modeling framework, which enables SBI for arbitrary summary statistics. {\sc SimBIG} accurately models galaxy clustering down to nonlinear scales and includes systematic effects in the observations. To demonstrate the framework, the authors adapted {\sc SimBIG} to the BOSS CMASS galaxy sample, and conducted SBI of $\Omega_m$, $\Omega_b$, $h$, $n_s$, and $\sigma_8$ from the galaxy power spectrum. Following in the footsteps of this study, we lead a similar analysis using the wavelet scattering transform (WST).

The WST, first introduced by \cite{Mallat2010, Mallat2012} from a data science perspective, defines a set of descriptive statistics well suited to the characterization of non-Gaussian fields resulting from nonlinear multiscale physics. In the past years, it has notably found interest in astrophysics for the modeling of the interstellar medium~\citep{Allys2019, Regaldo2020, Saydjari2021}, and in cosmology for the analysis of weak lensing data~\citep{Cheng2020, Cheng2021}, line-intensity mapping~\citep{Chung2022}, or the study of the epoch of reionization~\citep{Greig2022, Greig2023}. A 3D version of the WST first introduced in \cite{Eickenberg2017} for quantum chemistry applications enabled the extension of these statistics to the characterization of LSS~\citep{Eickenberg2022, Valogiannis2022a, Valogiannis2022b}. Fisher forecasts have shown that these statistics are much more constraining than the power spectrum on simulated matter density fields~\citep{Allys2020, Eickenberg2022, Valogiannis2022a}. Recently, \cite{Valogiannis2022b} showed that this is still the case on BOSS data in a Bayesian inference setting with strong assumptions on the likelihood function (including a Gaussian assumption). We now go further and conduct an SBI analysis of BOSS using the WST. We introduce a new variant of the WST statistics designed to account for the specificity of galaxy surveys, as well as further methodological developments to the {\sc SimBIG} inference pipeline for posterior estimation and validation.

The rest of this paper is organized as follows. In Sect.~\ref{sec:methodo}, we present our methodology. We specify our target observational data, and then introduce our inference pipeline and the WST statistics used as summary statistic. In Sect.~\ref{sec:results}, we present our results, which comprise a validation of our pipeline and the posterior distributions of the $\Lambda$CDM parameters. Finally, in Sect.~\ref{sec:conclusion}, we draw our conclusions and discuss perspectives. This paper also includes three appendices.
\section{Methodology}
\label{sec:methodo}

Our goal is to infer the $\Lambda$CDM cosmological parameters $\Omega_m$, $\Omega_b$, $h$, $n_s$, and $\sigma_8$ from the BOSS galaxy survey using SBI.
In Sect.~\ref{sec:target_data}, we introduce the target galaxy sample. We then give an overview of our forward model in Sect.~\ref{sec:inference_pipeline}. In Sect.~\ref{sec:summary_stats}, we present the WST summary statistics that we have designed for this work. Finally, in Sect.~\ref{sec:posterior_estimation}, we describe the SBI methodology.

\begin{figure*}
    \centering
    \includegraphics[scale=0.9]{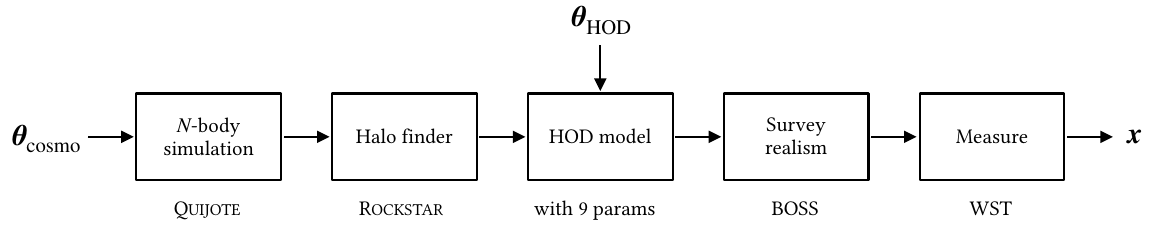}
    \caption{Diagram representing the different components of the {\sc SimBIG} forward model.}
    \label{fig:forward_model_diagram}
\end{figure*}

\subsection{Observations: BOSS CMASS Galaxies}
\label{sec:target_data}

In this paper, we analyze the CMASS galaxy sample of the BOSS Data Release 12. We limit this analysis to the southern galactic cap (SGC), which is further restricted in angular coordinates to ${\rm Dec} > -6~\deg$ and $-25 < {\rm RA} < 28~\deg$ and in redshift to $0.45 < z < 0.6$. In total, this catalog comprises 109,636 galaxies. We refer to \cite{Hahn2022a, Hahn2022b} for visuals of the sample.

\subsection{{\sc SimBIG} Forward Modeling Framework}
\label{sec:inference_pipeline}

We frame our inference problem in a Bayesian setting. Given a vector of measurements, $\ve{x}$, obtained from our galaxy sample, we aim to infer the posterior distribution of the vector of model parameters, $\ve{\theta}$. 
Bayes' theorem decomposes this posterior distribution as ${p(\ve{\theta}\,|\,\ve{x}) \propto p(\ve{x}\,|\,\ve{\theta}) p(\ve{\theta})}$, where $p(\ve{x}\,|\,\ve{\theta})$ is the likelihood function and $p(\ve{\theta})$ is the prior distribution.

With SBI~\citep[for a review, see][]{Cranmer2020}, the likelihood function is learned from simulated observations. This prevents us from using strong assumptions on its functional form. SBI only requires the ability to simulate a $\ve{x}$ for any given $\ve{\theta}$ through our forward model. For this analysis, we employ the {\sc SimBIG} forward model. We give below a general overview of this model and its parameters, and refer to \cite{Hahn2022a, Hahn2022b} for more details. Figure~\ref{fig:forward_model_diagram} gives a schematic representation of the components of the {\sc SimBIG} forward model. These include the following steps:
\begin{enumerate}
    \item An $N$-body simulation, that solves the dynamical evolution of a dark matter fluid in a $\Lambda$CDM cosmology parameterized by ${\ve{\theta}_{\rm cosmo} = (\Omega_m, \Omega_b, h, n_s, \sigma_8)}$. We use the existing high-resolution simulations of the {\sc Quijote} suite~\citep{Villaescusa2020}.
    \item A dark matter halo finder, which identifies halos from the dark matter particles. Here, we use {\sc Rockstar}~\citep{Behroozi2013}.
    \item A halo occupation distribution (HOD) framework designed to populate halos with galaxies in a phenomenological way. The HOD introduces another set of 9 HOD parameters $\ve{\theta}_{\rm HOD}$.
    \item The application of survey realism to mimic the BOSS CMASS SGC sample. This includes a reproduction of the survey geometry as well as observational systematics, including fiber collisions.
    \item The measurements of summary statistics from the final mock galaxy survey. This paper focuses on WST statistical measurements, described in detail in Sect.~\ref{sec:summary_stats}.
\end{enumerate}

\begin{table}
   \def\arraystretch{1.5}
   \centering
   \begin{tabular}{c|ccc}
       \hline
       \hline
       Cosmological & Prior & {\sc Quijote} & {\sc AbacusSummit} \vspace{-0.25cm}\\
       parameter &  & fiducial & fiducial\\
       \hline
       $\Omega_m$ & $\mathcal{U}(0.1, 0.5)$ & 0.3175 & 0.3152 \\
       $\Omega_b$ & $\mathcal{U}(0.03, 0.07)$ & 0.049 & 0.0493\\
       $h$ & $\mathcal{U}(0.5, 0.9)$ & 0.6711 & 0.6736\\
       $n_s$ & $\mathcal{U}(0.8, 1.2)$ & 0.9624 & 0.9649\\
       $\sigma_8$ & $\mathcal{U}(0.6, 1.0)$ & 0.834 & 0.8080\\
       \hline
       \hline
   \end{tabular}
   \caption{Prior distribution $p(\ve{\theta})$ and fiducial cosmologies~$\ve{\theta}_{\rm fid}$ used for the {\sc Quijote} and {\sc AbacusSummit} simulations.}
   \label{table:prior}
\end{table}

The target parameters $\ve{\theta}$ of the inference are $\ve{\theta}_{\rm cosmo}$ and $\ve{\theta}_{\rm HOD}$. 
We use the same prior distribution $p(\ve{\theta})$ as in \cite{Hahn2022b}. We recall in Table~\ref{table:prior} the priors on the cosmological parameters $\ve{\theta}_{\rm cosmo}$. These are conservative priors set by the range of parameters used to run the {\sc Quijote} simulations. We refer to \cite[][Table~1]{Hahn2022b} for a specification of the priors on $\ve{\theta}_{\rm HOD}$.

\subsection{Summary Statistics}
\label{sec:summary_stats}

In this work, we focus on WST statistics to capture the non-Gaussian information stemming from the nonlinear dynamical evolution of the LSS. We handcraft a new variant of WST statistics that leverage the symmetries of galaxy surveys. These include the power spectrum information as well as a quantification of interactions between scales.

\subsubsection{Wavelets}
\label{sec:wavelets}

Given an initial anisotropic 3D wavelet $\psi$ called the \textit{mother wavelet}, we build a family of wavelets $\{\psi_\lambda\}$ by dilation and rotation of $\psi$. 
With $(\ve{e}_x, \ve{e}_y, \ve{e}_z)$, the standard basis of $\mathbb{R}^3$, we introduce spherical coordinates $(r, \vartheta, \phi)$. 
$r$ is the radius, $\vartheta$ is the polar angle defined with respect to $\ve{e}_z$, and $\varphi$ is the azimuthal angle. We also introduce the unit radial vector ${\ve{e}_{\vartheta, \varphi} = \sin\vartheta\cos\varphi~\ve{e}_x + \sin\vartheta\sin\varphi~\ve{e}_y + \cos\vartheta~\ve{e}_z}$. Calling $j$ the index of dilation, the wavelet $\psi_\lambda$ with $\lambda=(j, \vartheta, \varphi)$ is a dilated (by a factor $2^j$) and rotated version of $\psi$, formally defined as follows:
\begin{equation}
    \psi_{j, \vartheta, \varphi}(\ve{r}) = 2^{-3j}\psi(T_{\vartheta, \varphi}2^{-j}\ve{r}),
\end{equation}
where $T_{\vartheta, \varphi}$ is a rotation matrix of $\mathrm{SO}(3)$ such that $T_{\vartheta, \varphi}\ve{e}_z = \ve{e}_{\vartheta, \varphi}$\footnote{Note that the $2^{-3j}$ prefactor used to normalize the wavelets has no impact on the rest of this work.}. Note that this only constrains $T_{\vartheta, \varphi}$ up to a rotation around $\ve{e}_{\vartheta, \varphi}$. 
However, since we later choose the mother wavelet $\psi$ to be axisymmetric, this definition will not depend on the particular choice of $T_{\vartheta, \varphi}$.

Inspired by \cite{Eickenberg2022}, we define the mother wavelet $\psi$ in Fourier space as the product of a radial component $R$ and an angular component $\alpha$, that is:
\begin{equation}
    \hat{\psi}(\ve{k}) = R(k)\alpha(\vartheta(\ve{k}), \varphi(\ve{k})).
\end{equation}
The radial component $R$ is chosen to be the Fourier transform of the spline (isotropic) wavelet introduced in \cite{Lanusse2012}.
This wavelet benefits from a good localization in physical space while being reasonably regular. Its definition is given in Appendix~\ref{app:wavelet}. Note that it depends on a cut-off frequency $k_c$ that delimits its support in Fourier space, with $R(k) = 0$ for $k > 2k_c$. For the angular component $\alpha$, we choose a Gaussian function that reads:
\begin{equation}
    \alpha(\vartheta) = \exp\left(-\frac{\vartheta^2}{2\sigma_\vartheta^2}\right),
\end{equation}
with $\sigma_\vartheta$ the angular width that parametrizes the localization of the function. Since $\alpha$ only depends on $\vartheta$, and not on $\varphi$, the resulting mother wavelet $\psi$ is axisymmetric. This feature will be helpful in the following to exploit the statistical invariance properties of redshift-space data.

Similarly to \cite{Eickenberg2022}, we employ dilation indices $j$ of the form $k/Q$ with $k \in \{0, \dots, J\times Q-1\}$. The parameter $J$ is the total number of octaves (doublings in scale) taken into account for the analysis, and $Q$ is a parameter called \textit{quality factor} that controls the number of scales per octave. The angular indices $\vartheta$ and $\varphi$ are chosen to be the 20 vertices of a regular dodecahedron\footnote{Or equivalently, the surface normals of a regular icosahedron. This choice corresponds to an invariant action space of the largest discrete non-planar transformation subgroup of SO(3), the icosahedral group.}. 
Initially orienting the regular dodecahedron such that one vertex coincides with the pole $\ve{e}_z$ ($\vartheta = 0$), we only keep the 10 vertices with $\vartheta \in [0, \pi/2]$.
This provides a good trade-off between computation speed and constraining power of the statistic.

\begin{figure*}
    \begin{subfigure}[b]{1.0\textwidth}
    \includegraphics[width=0.9\hsize]{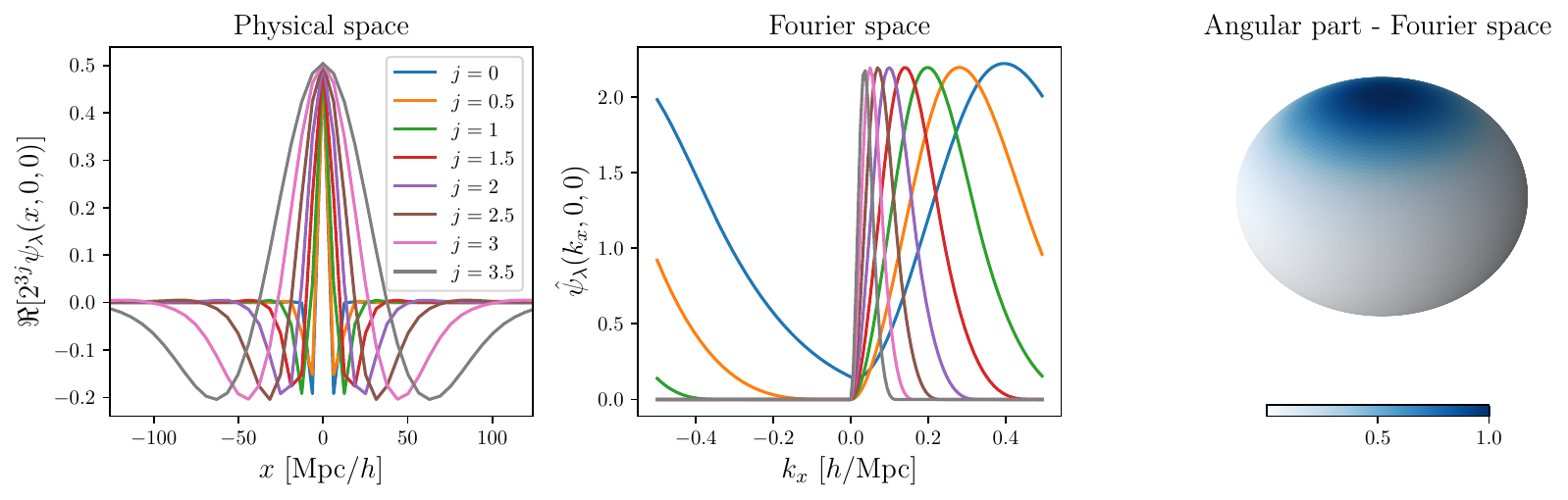}
    \caption{Radial (left and center) and angular (right) profiles for a subset of our 3D wavelets with fixed orientation. We show the radial profiles in both physical (left) and Fourier (center) space. The angular part is shown in Fourier space only. Our wavelets cover Fourier space with their passbands.}
    \end{subfigure}
    \begin{subfigure}[b]{1.0\textwidth}
    \includegraphics[width=0.85\hsize]{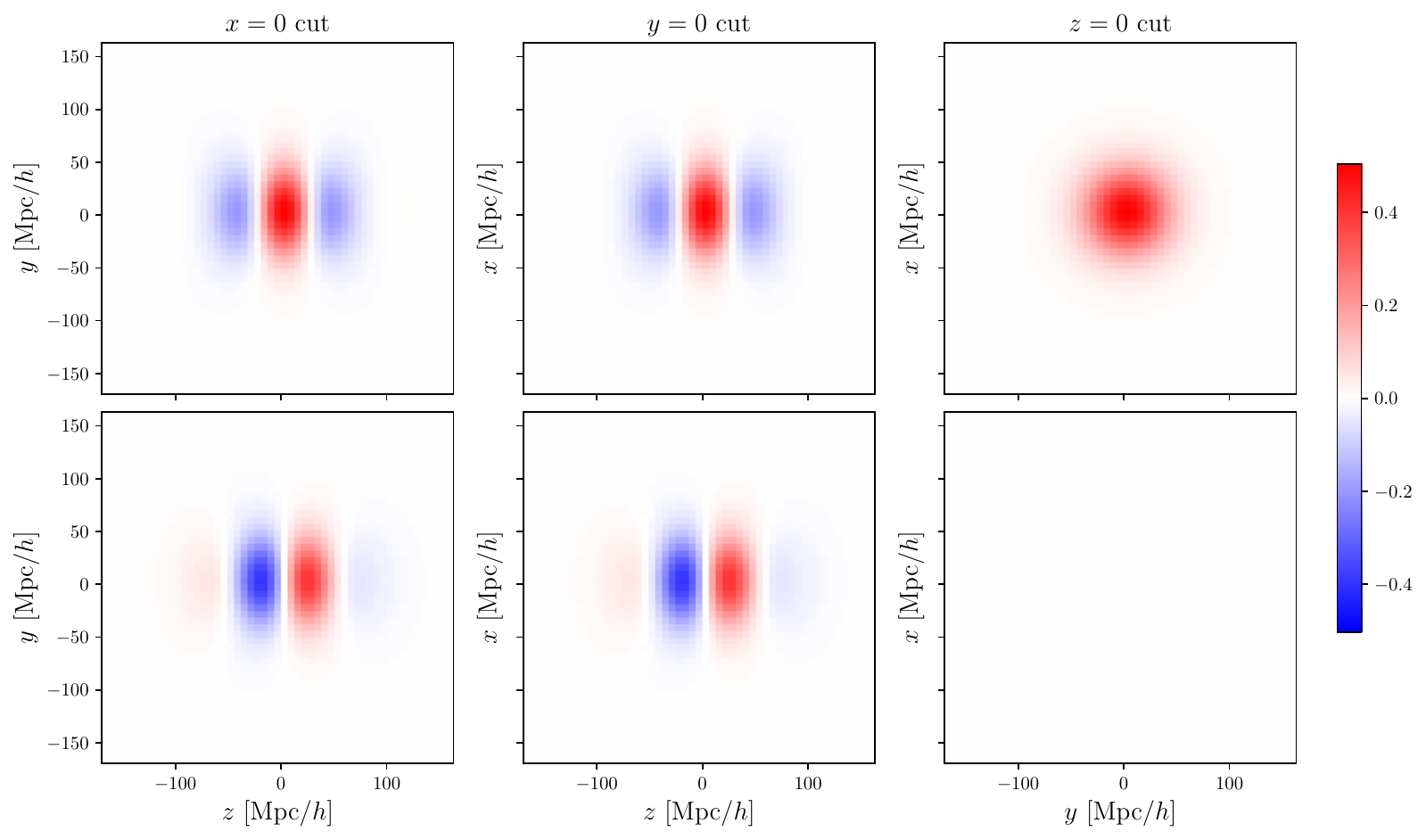}
    \caption{Planar cuts of the real (top row) and imaginary (bottom row) parts of $2^{3j}\psi_\lambda$ for $\lambda = (j, \vartheta, \varphi) = (3, 0, 0)$ (the wavelet is oriented along the $z$ axis). The planar cuts are shown for $x, y, z = 0$ from left to right, respectively.}
    \end{subfigure}
    \caption{Visualization of the 3D oriented wavelets designed for this work.}
    \label{fig:wavelets_visualization}
\end{figure*}

We set $k_c = 4\pi k_{\rm max}/ 3$, where $k_{\rm max}$ denotes the largest spatial frequency we want to probe. 
We also choose ${\sigma_\vartheta = \pi/4}$, as well as $J =6$ and $Q=2$. These parameters lead to a family of 120 wavelets. 
We visualize a subset of the wavelets in Fig.~\ref{fig:wavelets_visualization}.

\subsubsection{Wavelet Scattering Moments}
\label{sec:wst}

The definition of the WST coefficients involves operations that relate to the structure of convolutional neural networks, namely cascades of convolutions, pointwise nonlinearities, and average pooling operators.\footnote{One of the main motivations of the WST from the perspective of machine learning was to provide a mathematical framework to understand the properties of convolutional neural networks and the origin of their success for visual object classification problems~\citep{Mallat2016}.}
We define a generalized version of the wavelet scattering moments associated with a given real-valued field $F$ as follows:
\begin{align}
    S_0[p] &= \langle |F|^p \rangle, \\
    S_1[\lambda, p] &= \langle |F \ast \psi_\lambda|^p\rangle, \\
    S_2[\lambda_1, \lambda_2, p] &= \langle ||F \ast \psi_{\lambda_1}|\ast\psi_{\lambda_2}|^p\rangle,
\end{align}
where $\ast$ denotes the convolution operation, and $\langle \cdot \rangle$ is an averaging operator. These moments thus depend on wavelet indices $\lambda_i$ and exponents $p$. We refer to the $S_0$, $S_1$, and $S_2$ moments as the zeroth-, first-, and second-order moments, respectively. $S_0$ coefficients are moments estimates of the one-point function. $S_1$ coefficients measure moments of $x$ filtered at a given oriented scale $\lambda$.\footnote{$S_1$ coefficients with $p=2$ directly relate to the power spectrum~\citep{Mallat2012, Allys2020}.} Finally, $S_2$ coefficients quantify in an analogous way interactions between oriented scales $\lambda_1$ and $\lambda_2$. Note that $S_2$ coefficients with $j_2 \leq j_1$ are discarded since they usually become either non-informative or negligible~\citep{Bruna2013}.

A galaxy catalog takes the form of a point cloud, where the density field can be formally written as follows:
\begin{equation}
    n(\ve{r}) = \sum_i w_{c, i}\delta_{\rm D}(\ve{r} - \ve{r}_i),
\end{equation}
where $w_{c, i}$ are galaxy completeness weights.
The weights $w_{c, i}$ only play a role when measuring statistics on the BOSS data to compensate for imaging systematics or failures in measuring accurate redshifts~\citep[for a further discussion, see][]{Hahn2022b}. For the mock catalogs produced by our forward model, these are uniformly set to one.
We also introduce the field $\bar{n}(\ve{r})$ corresponding to the expected mean space density of galaxies~\citep[see][]{Feldman1994}. However, as it is usually done in the galaxy clustering analysis literature, and in order to mitigate edge effects due to the BOSS survey geometry in the computation of the WST coefficients, we focus on the density fluctuation field $F$ defined as:
\begin{equation}
\label{eq:fluctuation_field}
    F(\ve{r}) = \frac1{\sqrt{I}}\left[n(\ve{r}) - \alpha n_r(\ve{r})\right],
\end{equation}
where $I = \int \bar{n}^2(\ve{r})\dI \ve{r}$ is a normalization factor~\footnote{In practice, this term is estimated as described in \cite{Scoccimarro2015}.}, $n_r$ is a random catalog of more than 4 million objects with same survey geometry as the galaxy catalog, and $\alpha$ is a normalizing factor selected to make $F(\ve{r})$ have a spatial mean of zero.
Finally, we create a mesh representing $F(\ve{r})$ by interpolating the point cloud onto a 3D Cartesian grid using a triangular shape cloud interpolation scheme from {\sc nbodykit}~\citep{Hand2018}.
The grid spacing is chosen to be $\lambda_{\rm max} / 2$, where $\lambda_{\rm max} = 2\pi / k_{\rm max}$, which thus satisfies the Nyquist criterion. The wavelets are also discretized on the same grid. Similarly to \cite{Hahn2022b}, we restrict our analysis to spatial frequencies lower than $k_{\rm max} = 0.5~h/\textrm{Mpc}$.

%Note that, for the power spectrum statistics, it is usual to add to Eq.~\eqref{eq:fluctuation_field} a global weighting field designed to minimize the variance of the power spectrum estimator~\citep[see][]{Feldman1994}. 
%%We could have proceeded in a similar fashion for the WST statistics to improve the variance of their estimators. However, this investigation did not seem straightforward to us, and is left for future work. 
%Similar considerations may be possible for WST statistics or a subset thereof. We leave this investigation for future work.

\subsubsection{Reduced Moments}

With simple invariance assumptions on $F$ that reflect the cosmological principle, it is possible to define so-called \textit{reduced} WST moments which lead to a set of descriptive statistics for $F$ of lower dimension, reduced variance, and which are often easier to interpret~\citep{Allys2019, Regaldo2020}. We first normalize the WST coefficients as follows:
\begin{align}
    \bar{S}_0[p] &= \log\left(S_0[p]\right), \\
    \bar{S}_1[\lambda, p] &= \log\left(\frac{S_1[\lambda, p]}{S_0[p]}\right), \\
    \bar{S}_2[\lambda_1, \lambda_2, p] &= \log\left(\frac{S_2[\lambda_1, \lambda_2, p]}{S_1[\lambda_1, p]}\right).
\end{align}
This procedure helps to disentangle the information measured by these different coefficients\footnote{This makes $\bar{S}_1$ coefficients invariant to global scaling of $F$, and $\bar{S}_2$ coefficients invariant to linear filtering of $F$ with top-hat passbands including that of $\psi_{\lambda_1}$.}, and evens their magnitudes.

\begin{figure}
    \centering
    \includegraphics{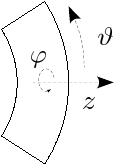}
    \caption{Definition of the spherical coordinates with respect to the survey geometry.
    The redshift axis is $z$.
    The elevation angle $\vartheta$ measures deviation from $z$.
    The azimuthal angle $\varphi$ measures rotation around $z$.
    The rounded rectangle represents a cut through the survey geometry.}
    \label{fig:frame}
\end{figure}

Let us first mention that, by defining globally averaged WST coefficients in Sect.~\ref{sec:wst}, we had already ignored the potential dependency of the coefficients with cosmological redshift, as if $F$ were statistically homogeneous. In light of the cosmological principle, this assumption is only rigorously justified on surfaces of constant cosmological redshift. However, given the narrow redshift range of our data ($0.45 < z < 0.6$), we consider this approximation reasonable and leave the investigation of local WST statistics for cosmological inference to further work. Similarly, we expect statistical invariance of $F$ with respect to rotations around the redshift axis. By neglecting the curvature of the survey domain and choosing $\ve{e}_z$ for the redshift axis, rotational invariance around $\ve{e}_z$ corresponds to an invariance with respect to the azimuthal angle $\varphi$ (see Fig.~\ref{fig:frame} for an illustration).

We leverage this rotational invariance by introducing the following first- and second-order reduced moments:
\begin{align}
    \hat{S}_1[j, \vartheta, p] &= \langle \bar{S}_1[\lambda, p]\rangle_\varphi, \\
    \hat{S}_2[j_1, \vartheta_1, j_2, \vartheta_2, \delta\varphi, p] &= \langle \bar{S}_2[\lambda_1, \lambda_2, p]\rangle_{|\varphi_1 -\varphi_2| = \delta\varphi},
\end{align}
where $\langle \cdot \rangle_{\varphi}$ and $\langle \cdot \rangle_{|\varphi_1 -\varphi_2| = \delta\varphi}$ stand for averages over $\varphi$ and the set of $(\varphi_1, \varphi_2)$ pairs with fixed distance ${|\varphi_1 -\varphi_2| = \delta\varphi}$, respectively. The dependence of these coefficients on the angular variables quantifies the impact of redshift-space distortions.

In the following, we define the vector of summary statistics $\ve{x}$ as the concatenation of zeroth-, first-, and second-order reduced coefficients, that is 3,675 coefficients in total. On a secondary level, we will also include in that vector the total number of objects in the input catalog $n_{\rm gal}$ following \cite{Hahn2022b}, as well as the shot noise power $P_{\rm shot}$ computed as in \cite{Scoccimarro2015}. The impact of shot noise on the WST coefficients has not been studied yet, this is the reason why we include this $P_{\rm shot}$ coefficient in the (idealistic) idea that this impact will be directly learned from the data. Future works will address this issue in a more straightforward way.

\subsection{Simulation-Based Inference}
\label{sec:posterior_estimation}

We use WST measurements from 20,000 mock galaxy catalogs constructed using the {\sc SimBIG} forward model. 
Formally, these measurements correspond to samples $(\ve{\theta}_1, \ve{x}_1), \dots, (\ve{\theta}_n, \ve{x}_n) \,\sim\, p(\ve{\theta}, \ve{x})$ drawn from the joint distribution, where $n = 20,000$. 
With SBI, we estimate the posterior, $p(\ve{\theta}\,|\,\ve{x})$
solely based on these samples.

To do that, we train a conditional normalizing flow~\citep[for reviews on normalizing flows, see][]{Kobyzev2021, Papamakarios2022} using the Python package {\sc sbi}~\citep{Tejero-Cantero2020}. We use a masked autoregressive flow (MAF) architecture~\citep{Papamakarios2017}, which consists of a sequence of autoregressive affine transforms whose coefficients are given by the output of a multilayer perceptron (MLP) neural network. With $q_\phi(\ve{\theta}\,|\,\ve{x})$, a MAF architecture with unknown parameters $\phi$, the training procedure consists in minimizing the following loss function:
\begin{align}
    \mathcal{L}(\phi) &= D_{\rm KL}\left[p(\ve{\theta}, \ve{x}) || q_\phi(\ve{\theta}\,|\,\ve{x})p(\ve{x})\right], \\
    &\approx \sum_{(\ve{\theta}_i, \ve{x}_i) \in T} \log p(\ve{\theta}_i\,|\,\ve{x}_i) - \log q_\phi(\ve{\theta}_i\,|\,\ve{x}_i),
\end{align}
where $D_{\rm KL}$ is the Kullback-Leibler divergence, and $T$ is our training set, which is selected to be a subsample of the full data set $\{(\ve{\theta}_i, \ve{x}_i)\}_{1 \leq i \leq n}$.\footnote{For further details on this derivation, we refer the reader to \cite{Hahn2022a}.} This minimization is equivalent to the maximization of the training score:
\begin{equation}
    \mathcal{S}(\phi) = \sum_{i \in T} \log q_\phi(\ve{\theta}_i\,|\,\ve{x}_i),
\end{equation}
which we use in practice. 

We perform this optimization using {\sc Adam}~\citep{Kingma2015}. It is regularized with an early-stopping procedure\footnote{We stop the optimization when the validation score does not increase after 20 consecutive epochs.}, with a train/validation split of 90\%/10\%. Hyperparameters, including the number of affine transforms used in the MAF, number of blocks and hidden units of the MLP, dropout probability of the MLP dropout layers (for further regularization), learning rate, and batch size, are all tuned to achieve the best validation score. In practice, we explore the hyperparameter space with a random search monitored by {\sc WandB}~\citep{Biewald2020} for $\sim 2,000$ models. We provide details on the ``sweep" configuration used for this random search in Appendix~\ref{app:wandb}.

To further increase the score $\mathcal{S}(\phi)$, we define an ensemble model from a linear mixture of the ten best models previously optimized as follows~\citep{Lakshminarayanan2016, Alsing2019}:
\begin{equation} \label{eq:ensembling}
    q_{\rm ens}(\ve{\theta}\,|\,\ve{x}) = \sum_{i=1}^{10}w_i q_{\phi_i}(\ve{\theta}\,|\,\ve{x}).
\end{equation}
The weights $\{w_i\}$ are positive numbers such that $\sum_i^{10} w_i = 1$ (thus ensuring correct normalization of $q_{\rm ens}$). For this work, a uniform weighting, $w_i = 1/10$, has been sufficient to significantly improve the training and validation scores. Choosing a uniform weighting is suggested only after carefully narrowing down the hyperparameter space through successive random searches~(see Appendix~\ref{app:wandb} for further details). Note that a more general strategy to optimize these weights without such costly successive random searches has been very recently investigated in this context~\citep{Yao2023}. This technique will be used in future work.

Let us finally mention that we also have explored a more expressive alternative to the MAF, namely a neural spline flow~\citep{Durkan2019}. However, we faced overfitting issues that we were not able to resolve with standard regularization approaches. Further exploration is left for future work.

\section{Results and discussion}
\label{sec:results}

We present our results in this section. In Sect.~\ref{sec:validation}, we validate our inference pipeline, and study its robustness to the change of forward models following \cite{Hahn2022b}. In Sect.~\ref{sec:boss_analysis}, we present the results of our inference on the BOSS data.

\subsection{Validation}
\label{sec:validation}

We validate our posterior estimation in two stages. 
First, we assess whether $q_{\rm ens}(\ve{\theta}\,|\,\ve{x})$ accurately estimates the true posterior distribution $p(\ve{\theta}\,|\,\ve{x})$ using simulation-based calibration~\citep[SBC,][]{Talts2018}.
Second, we conduct the {\sc SimBIG} ``mock challenge'', where we assess the robustness of our inference pipeline using a suite of test simulations. \\

\begin{figure*}
    \centering
    \includegraphics[width=\textwidth]{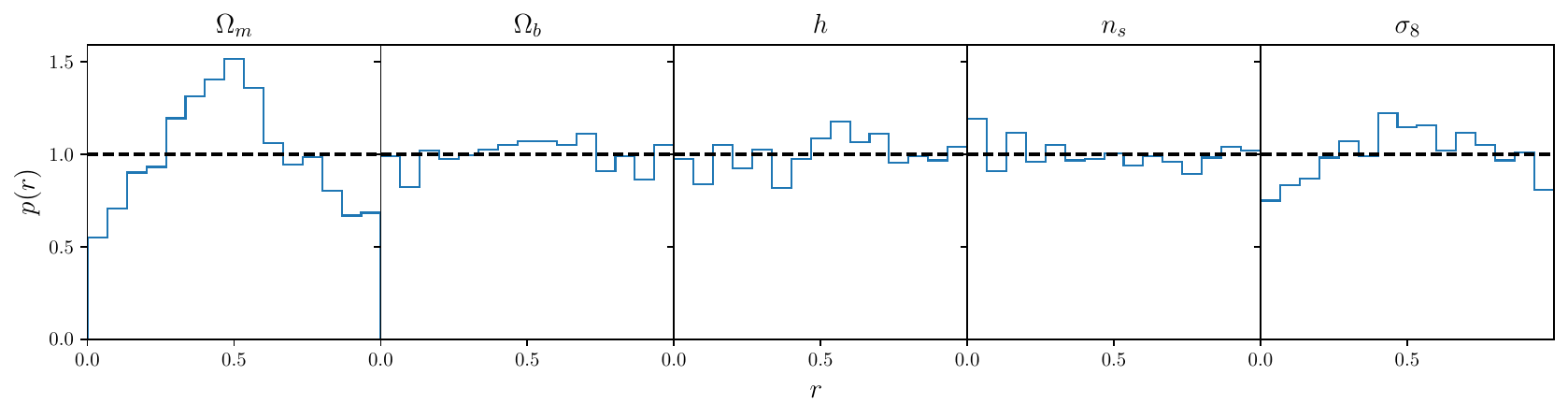}
    \caption{Posterior accuracy test using SBC. We present the distributions of the rank statistics 
    (Sect.~\ref{sec:validation}) for each of the cosmological parameters.
    These are derived from the validation set used during posterior estimation.
    The rank statistics should be uniformly distributed for an ideal posterior estimate (black dashed).
    Our results indicate accurate estimates for $\Omega_b$, $h$, and $n_s$ but underconfidence for $\Omega_m$ and $\sigma_8$. 
    }
    \label{fig:sbc_rank}
\end{figure*}

\paragraph{Posterior Accuracy Test.} We perform simulation-based calibration (SBC) to check the validity of our posterior estimation following \cite{Talts2018}. For each sample $(\tilde{\ve{\theta}}, \tilde{\ve{x}})$ of the validation set used during the training procedure\footnote{The limited amount of data prevents us from using an independent test set. However, since this validation set only served for the early-stopping procedure mentioned in Sect.~\ref{sec:posterior_estimation}, we do not expect significantly different results even with a separate test set.} (Sect.~\ref{sec:posterior_estimation}), we generate $m = 2,500$ independent samples from the posterior distribution $q_{\rm ens}(\ve{\theta}\,|\,\tilde{\ve{x}})$, and then compute the (normalized) rank of each component of the vector $\tilde{\ve{\theta}}$ among the corresponding components of these samples.

We present the distributions of these ranks for each cosmological parameter in Fig.~\ref{fig:sbc_rank}.
For an ideal posterior estimation, the rank statistics should be uniformly distributed over $[0,1]$. For $\Omega_b$, $h$, and $n_s$, these are close to uniform, which indicates accurate predictions for these parameters. However, for $\Omega_m$ and $\sigma_8$, we observe significant $\cap$-like patterns, which suggest underconfident predictions on these parameters. 
This is likely due to the limited size of the training set and underlines the challenges of posterior estimation in this high-dimensional context. 
Nevertheless, we consider that underconfidence does not undermine our analysis, but rather suggests that the constraining power of the WST statistics may be underexploited.
Let us finally point out that the ensembling procedure (Eq.~\ref{eq:ensembling}) plays a significant role in avoiding overconfidence on $\Omega_b$, $h$, and $n_s$. Indeed, separate SBC analyses of the ten best models have shown a slight tendency for overconfidence on these parameters, which has been corrected by the ensembling. \\

\begin{figure*}
    \centering
    \includegraphics[width=\textwidth]{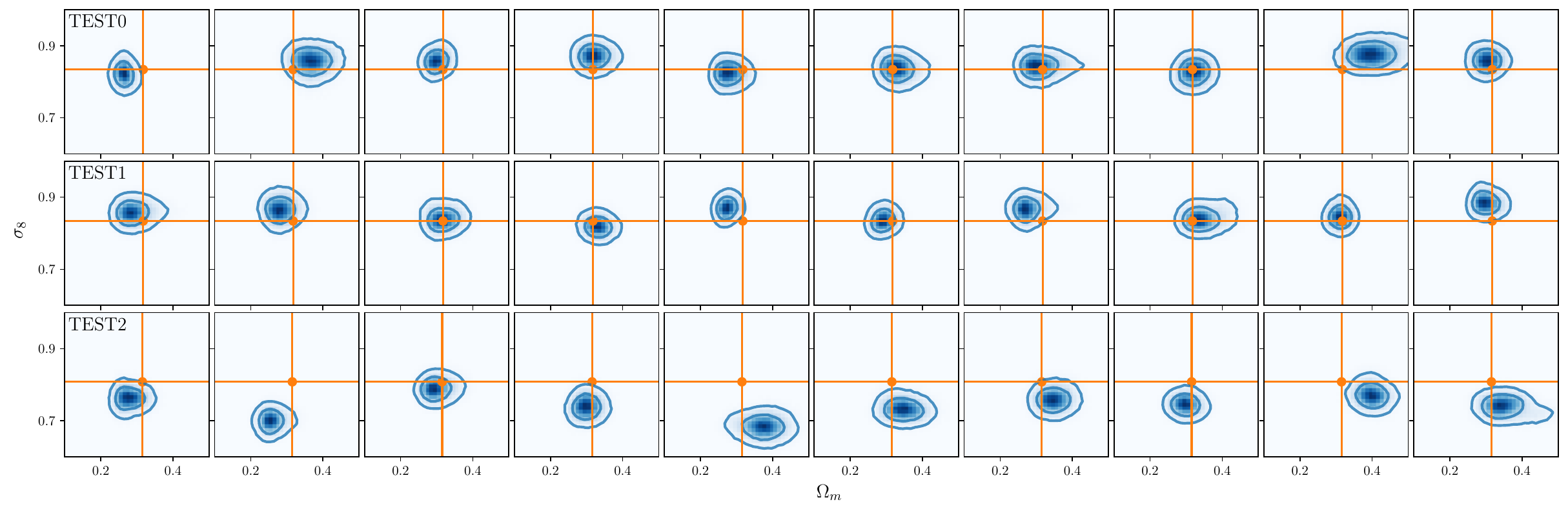}
    \caption{Posterior distributions for $(\Omega_m, \sigma_8)$ inferred from WST measurements of 10 randomly selected test simulations from: $\mathtt{TEST0}$ (top), $\mathtt{TEST2}$ (middle), and $\mathtt{TEST2}$ (bottom).
    The true parameters are marked in orange, and the contours represent the 68 and 95 percentiles.
    For $\mathtt{TEST0}$ and $\mathtt{TEST1}$, the true parameters lie expectedly within the percentiles of the posteriors. 
    However, for $\mathtt{TEST2}$, the true $\sigma_8$ tend to lie systematically higher than the inferred posteriors. 
    }
    \label{fig:test_posteriors}
\end{figure*}

\begin{figure*}
    \centering
    \includegraphics[width=\textwidth]{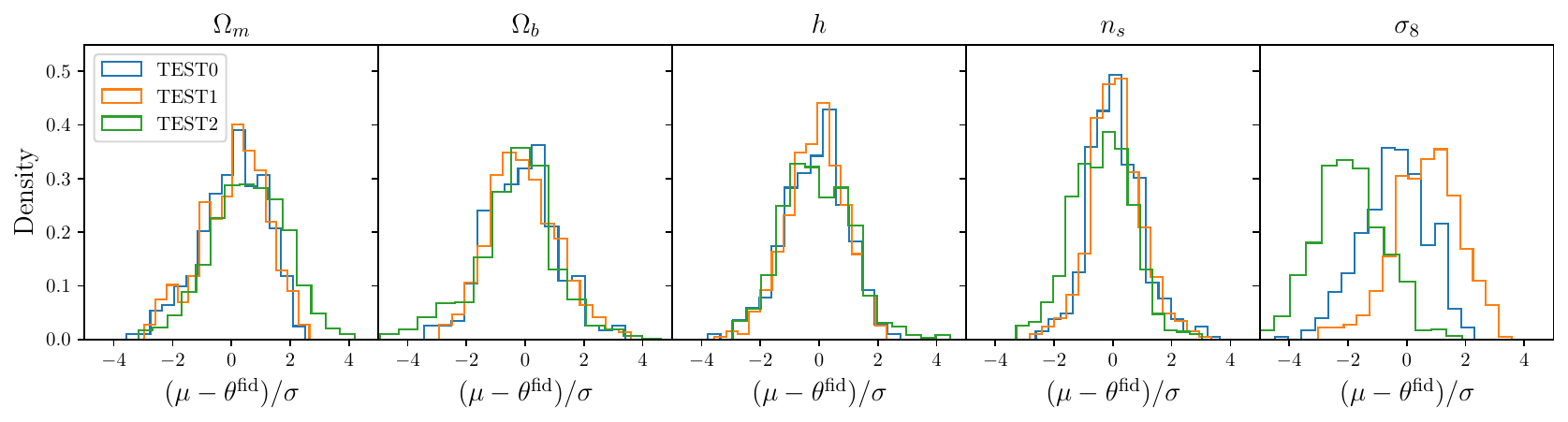}
    \caption{Distributions of the differences between the posterior mean $\mu$ and the true parameter $\theta^{\rm fid}$ normalized by the posterior standard deviation $\sigma$, for each cosmological parameter and test set.
    The distributions agree across the test sets for $\Omega_m$, $\Omega_b$, $h$, and $n_s$, but not for $\sigma_8$. This indicates that our predictions for $\sigma_8$ are not robust to variations of the forward model.
    }
    \label{fig:complike}
\end{figure*}

\paragraph{{\sc SimBIG} Mock Challenge.} Next, we assess the generalization capability and robustness of our inference pipeline by applying it to the {\sc SimBIG} mock challenge data introduced in~\cite{Hahn2022b}. This data consists of the following test sets:
\begin{itemize}
\item $\mathtt{TEST0}$: 500 mock catalogs built from the {\sc SimBIG} forward model described in Sect.~\ref{sec:inference_pipeline}, but using a set of 100 {\sc Quijote} simulations at a fiducial cosmology (see Table~\ref{table:prior}) and with HOD parameters drawn from a narrower distribution than the prior distribution.
\item $\mathtt{TEST1}$: 500 mock catalogs built from a similar forward model with the same fiducial {\sc Quijote} simulations, but a different friend-of-friend halo finder~\citep{Davis1985} and a simplified HOD model involving 6 parameters instead of 9.
\item $\mathtt{TEST2}$: 1,000 mock catalogs built from the {\sc AbacusSummit} $N$-body simulations~\citep{Maksimova2021, Garrison2021} and the {\sc CompaSO} halo finder~\citep{Hadzhiyska2022}. The simulations are constructed at a slightly different fiducial cosmology from {\sc Quijote} (Table~\ref{table:prior}). Otherwise, we use the same HOD model as the {\sc SimBIG} forward model.
\end{itemize}
Since $\mathtt{TEST0}$ uses the {\sc SimBIG} forward model, it serves as a means to quantify the ability of our pipeline to generalize to new unseen data. On the other hand, $\mathtt{TEST1}$ and $\mathtt{TEST2}$, which use different forward models, help us to quantify the robustness of our pipeline.

We show in Fig.~\ref{fig:test_posteriors}, for each of the test sets, examples of posterior distributions inferred from randomly selected test simulations. 
The top, center, and bottom rows present simulations from the $\mathtt{TEST0}$, $\mathtt{TEST1}$, and $\mathtt{TEST2}$, respectively.
We focus on $(\Omega_m, \sigma_8)$ since these are the parameters mostly significantly constrained by galaxy clustering. 
On the $\mathtt{TEST0}$ and $\mathtt{TEST1}$ examples, the true parameters are well within the 95\% contours on almost all examples. 
The $\mathtt{TEST0}$ results visually demonstrate the ability of our pipeline to generalize to new unseen data, while the $\mathtt{TEST1}$ results indicate first elements of robustness to the change of the forward model.
However, in comparison, the $\mathtt{TEST2}$ results show contours that are almost always below the true parameters. This indicates a bias in the $\sigma_8$ predictions, suggesting that our WST analysis is not fully robust to changes in the forward model.

We now complement the visual validation results with a quantitative assessment. 
For each test simulation, we compute the posterior mean $\mu$ and standard deviation $\sigma$, and focus on the difference between $\mu$ and the true parameter $\theta^{\rm fid}$ in units of $\sigma$ for each cosmological parameter. Although we do not expect the posterior mean $\mu$ to be an unbiased estimate of the true parameter $\theta^{\rm fid}$, for a robust pipeline, we do expect a form of statistical consistency of these estimates across the test sets.
We show in Fig.~\ref{fig:complike} the distributions of $(\mu - \theta^{\rm fid})/\sigma$ for each cosmological parameter. We consider that our pipeline is robust if these distributions are invariant to the change of forward model.\footnote{One interpretation of these distributions is that they are the likelihoods of a compression of the WST statistics. Thus, the comparison examines whether the compressed WST likelihoods are robust to change in the forward model.}
These results indicate robustness for all parameters except $\sigma_8$. The $\sigma_8$ histograms confirm what we find in Fig.~\ref{fig:test_posteriors}: our posterior estimate systematically infers a lower value of $\sigma_8$ for $\mathtt{TEST2}$.
We also find that our posterior estimate systematically infers in average a higher value of $\sigma_8$ for $\mathtt{TEST1}$. 
This demonstrates that the $\sigma_8$ posteriors are not robust to the alternative forward models considered in the {\sc SimBIG} mock challenge. 
The results of the mock challenge imply that: 1)~the change of forward model has a significant impact on galaxy clustering, 2)~the WST is sensitive to this impact, and  3)~our inference pipeline erroneously attributes this to variations of $\sigma_8$. \\

\begin{figure}
    \centering
    \includegraphics[scale=0.6]{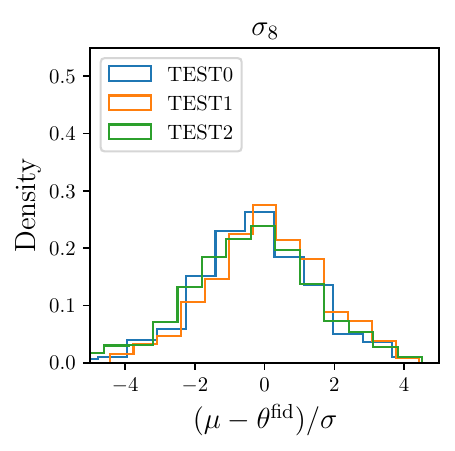}
    \caption{Same as Fig.~\ref{fig:complike} but using summary statistics where WST coefficients with $j < j_{\rm min} = 2$ are discarded.  This removes galaxy clustering information for scales smaller than ${k_{\rm max} \approx 0.3\,h/{\rm Mpc}}$.
    We focus on $\sigma_8$.
    By imposing a stringent scale cut we improve the robustness of our WST analysis.}
    \label{fig:tests_Om_s8_jmin_2}
\end{figure}

\paragraph{Addressing Non-Robustness.} We make a first attempt to address the robustness issues on $\sigma_8$. 
Since we expect galaxy clustering to better agree across different forward models on larger scales closer to the linear regime, we introduce stringent scale cuts on the WST statistic.
Introducing the scale index $j_{\rm min}$, we define new summary statistics where WST coefficients involving wavelets with $j < j_{\rm min}$ are discarded. 
We progressively increase $j_{\rm min}$ and repeat our validation test. We are able to resolve the robustness issues raised by the mock challenge for $j_{\rm min} = 2$. 
This discards any information from modes with $k \gtrsim 0.3~h/\text{Mpc}$ (see Fig.~\ref{fig:wavelets_visualization}~(a)). 
Figure~\ref{fig:tests_Om_s8_jmin_2} shows that the  $(\mu - \theta^{\rm fid})/\sigma$ distributions obtained for $j_{\rm min} = 2$ now agree for $\sigma_8$.
Interestingly, the equivalent distributions for the other cosmological parameters are not impacted by this scale cut.
This suggests that small-scale clustering was mostly the origin of the point 3) raised above.

\subsection{BOSS WST Analysis}
\label{sec:boss_analysis}

\begin{figure*}
    \includegraphics[width=0.49\textwidth]{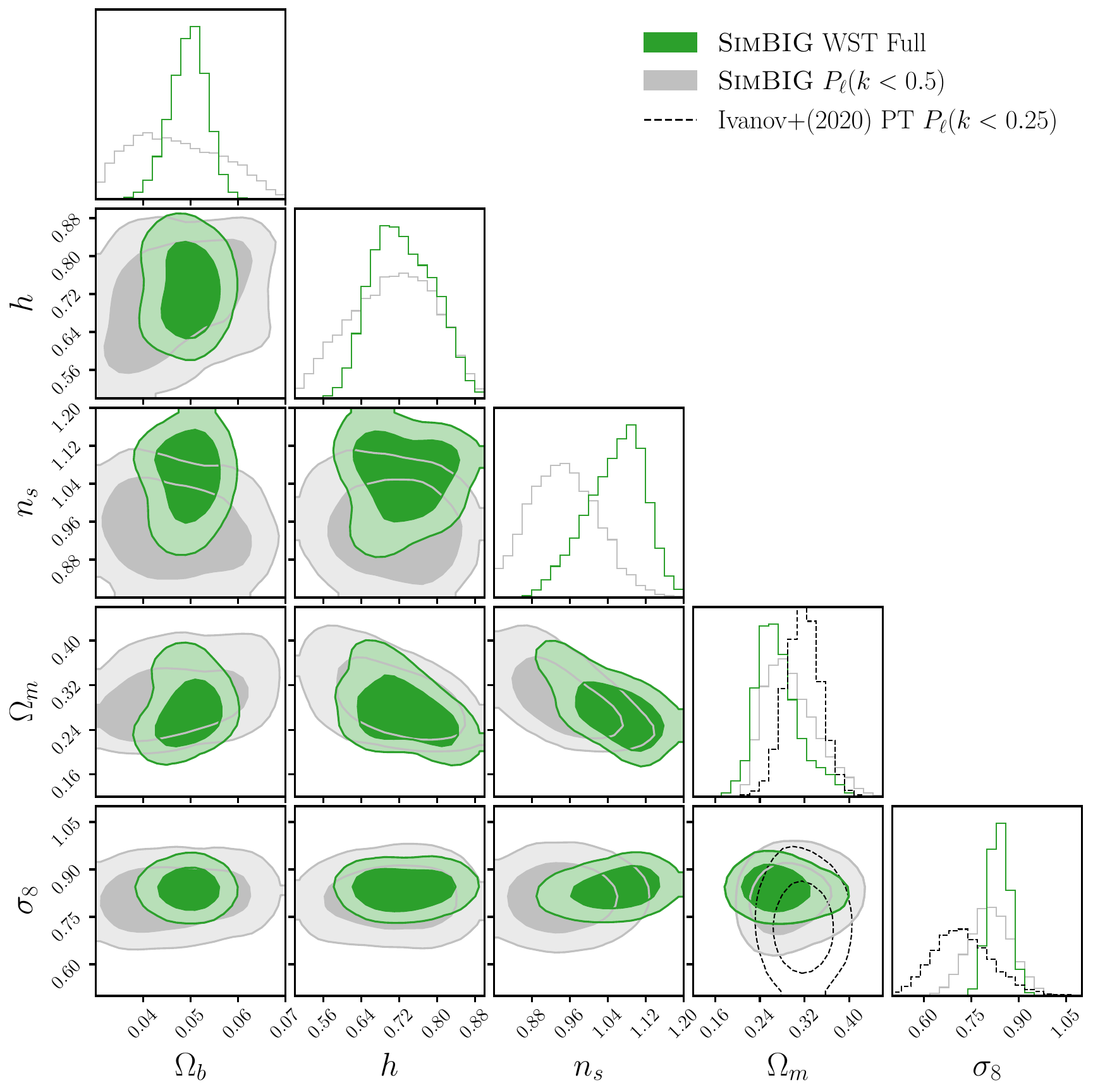}
    \includegraphics[width=0.49\textwidth]{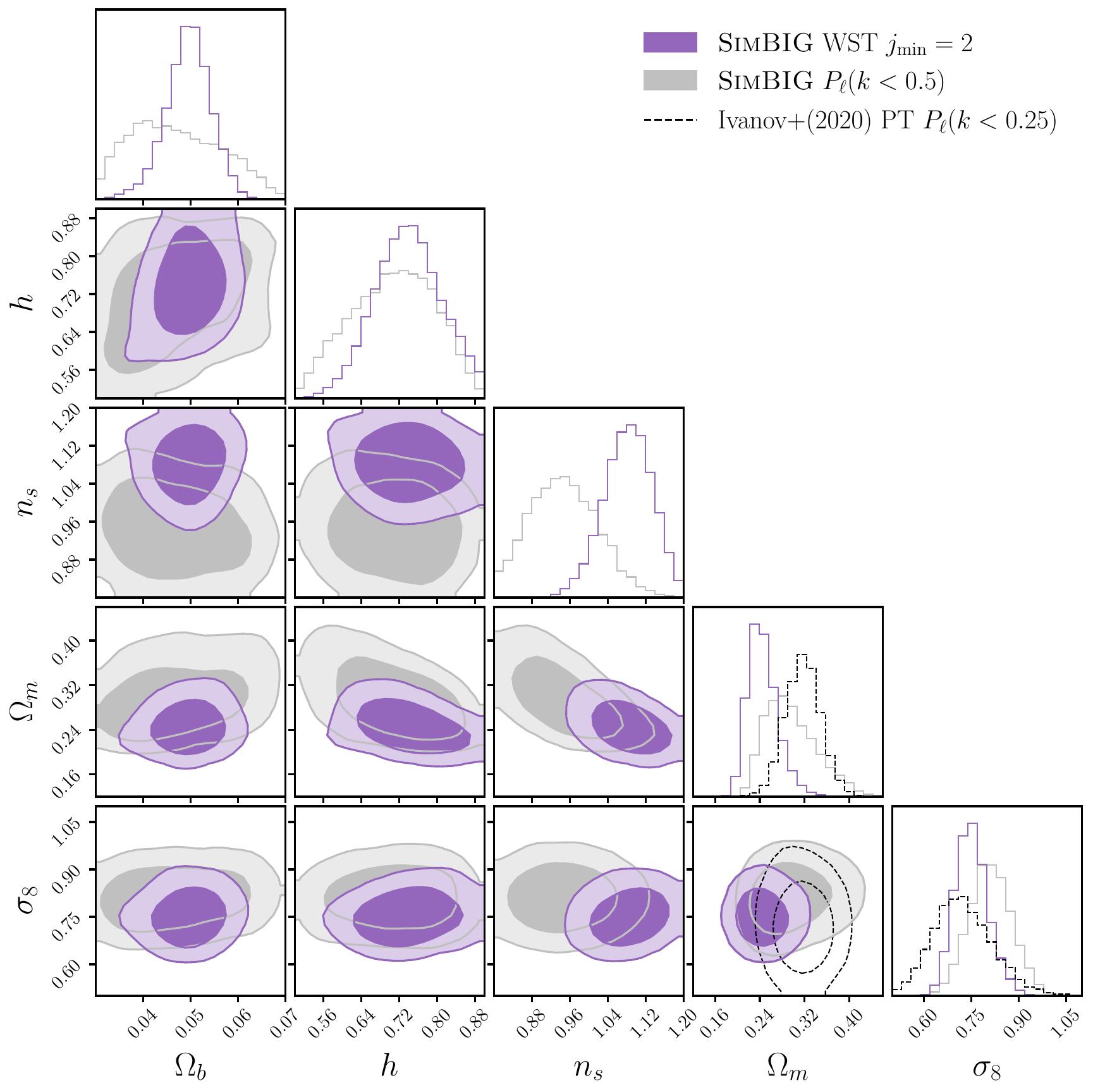}
    \caption{Posterior distributions for the $\Lambda$CDM cosmological parameters 
    inferred from the BOSS CMASS SGC sample using WST with {\sc SimBIG}. 
    On the left, we present constraints using the full WST statistics, which probe scales down to ${k_{\rm max} = 0.5~h/\text{Mpc}}$.
    On the right we present constraints using the $j_{\rm min}=2$ WST statistics, where modes with $k \gtrsim 0.3\,h/{\rm Mpc}$ are discarded.
    The contours represent the 68 and 95 percentiles. 
    We also show for comparison posteriors from the {\sc SimBIG} $P_\ell$ analysis~\citep[grey;][]{Hahn2022a, Hahn2022b} and the \cite{Ivanov2020} PT based $P_\ell$ analysis (black dashed). 
    }
    \label{fig:obs_posterior}
\end{figure*}

\begin{table*}
    \def\arraystretch{1.5}
    \centering
    \begin{tabular}{c|c|ccccc}
        \hline
        \hline
        Analysis & Quantity & $\Omega_m$ & $\Omega_b$ & $h$ & $n_s$ & $\sigma_8$ \\
        \hline
        \hline
        \multirow{2}{10em}{\centering {\sc SimBIG}~WST (full)} & mean $\pm \sigma$ & $0.28 \pm 0.04$ & $0.050 \pm 0.004$ & $0.73 \pm 0.06$ & $1.05 \pm 0.06$ & $0.84 \pm 0.03$ \\
 & median \& 68\% CI & $0.27_{-0.03}^{+0.04}$ & $0.051_{-0.004}^{+0.004}$ & $0.73_{-0.06}^{+0.06}$ & $1.06_{-0.07}^{+0.06}$ & $0.83_{-0.03}^{+0.03}$ \\ 
        \hline
        \multirow{2}{10em}{\centering {\sc SimBIG}~WST ${j_{\rm min} = 2}$} & mean $\pm \sigma$ & $0.25 \pm 0.03$ & $0.050 \pm 0.005$ & $0.74 \pm 0.07$ & $1.08 \pm 0.05$ & $0.75 \pm 0.05$ \\
 & median \& 68\% CI & $0.24_{-0.02}^{+0.03}$ & $0.050_{-0.005}^{+0.004}$ & $0.74_{-0.07}^{+0.07}$ & $1.08_{-0.05}^{+0.05}$ & $0.75_{-0.05}^{+0.05}$ \\ 
    \hline
    \multirow{2}{10em}{\centering {\sc SimBIG} ${P_\ell(k < 0.5\,h/{\rm Mpc})}$} & mean $\pm \sigma$ & $0.29 \pm 0.04$ & $0.046 \pm 0.008$ & $0.70 \pm 0.09$ & $0.94 \pm 0.07$ & $0.81 \pm 0.07$ \\
 & median \& 68\% CI & $0.28_{-0.04}^{+0.05}$ & $0.046_{-0.009}^{+0.009}$ & $0.70_{-0.10}^{+0.11}$ & $0.93_{-0.07}^{+0.08}$ & $0.81_{-0.08}^{+0.07}$ \\ 
    \hline
    \multirow{2}{10em}{\centering PT ${P_\ell(k < 0.25\,h/{\rm Mpc})}$} & mean $\pm \sigma$ & $0.32 \pm 0.03$ & $0.052 \pm 0.007$ & $0.68 \pm 0.06$ & $0.96 \pm 0.05$ & $0.73 \pm 0.10$ \\
         & median \& 68\% CI & $0.32_{-0.03}^{+0.03}$ & $0.052_{-0.007}^{+0.007}$ & $0.68_{-0.06}^{+0.06}$ & $0.95_{-0.05}^{+0.07}$ & $0.72_{-0.09}^{+0.10}$ \\ 
         \hline
         \hline
    \end{tabular}
    \caption{Constraints on the $\Lambda$CDM cosmological parameters inferred from the BOSS CMASS SGC using the WST with {\sc SimBIG}. For each parameter, we report the posterior mean and standard deviation, as well as the 68\% credible interval centered on the median. We also show for comparison the constraints from the {\sc SimBIG} $P_\ell$ analysis~\citep{Hahn2022a, Hahn2022b} and the \cite{Ivanov2020} PT-based $P_\ell$ analysis.}
    \label{table:obs_stats}
\end{table*}

We now analyze the BOSS CMASS galaxy sample with our inference pipeline.
We show in Fig.~\ref{fig:obs_posterior} the posterior distribution of the cosmological parameters obtained from both the full WST (left) 
and $j_{\rm min} = 2$ WST statistics (right). 
For the $j_{\rm min} = 2$ results, we use an ensemble of 50 models instead of 10 (see Sect.~\ref{sec:posterior_estimation}) for reasons explained below.
In both cases, we compare our results to two distinct power spectrum analyses on the same data: the {\sc SimBIG} $P_\ell$ analysis~\citep{Hahn2022a, Hahn2022b} and the \cite{Ivanov2020} $P_\ell$ analysis based on PT.
The {\sc SimBIG} and \cite{Ivanov2020} $P_\ell$ analyses include scales down to $k_{\rm max} = 0.5$ and $0.25~h/{\rm Mpc}$, respectively. 
We also show for reference in Fig.~\ref{fig:obs_full_posterior} and \ref{fig:obs_full_posterior_jmin_2} equivalent plots that include the HOD parameters.
We finally summarize in Table~\ref{table:obs_stats} all the posterior mean, standard deviation, median, and 68\% credible interval. 
The $j_{\rm min} = 2$ WST analysis yields significantly tighter constraints on almost all parameters while being consistent with the $P_\ell$ analyses. 
Compared to the standard PT-based ${P_\ell(k < 0.25\,h/{\rm Mpc})}$ analysis, we tighten the 68\% CI constraints for $\Omega_m$, $\Omega_b$, $h$, $n_s$, and $\sigma_8$ by a factor $1.2$, $1.5$, $0.8$, $1.3$, and $1.8$, respectively. These are clear improvements for all parameters except $h$. The deterioration of the constraints for $h$ could be explained by the smaller BOSS CMASS sample size that we use compared to \cite{Ivanov2020}.

However, we raise a notable concern on these results. 
For both the $j_{\rm min} = 0$ and $j_{\rm min} = 2$ analyses, we find that the separate models included in the ensemble (Eq.~\ref{eq:ensembling}) infer highly variable posteriors from one another (see Fig.~\ref{fig:obs_posterior_modelmisspec_jmin_2}).
This is the case even for models with similar validation scores and that pass the posterior accuracy test.
This could be a sign of model misspecfication. In other words, the WST statistics of the BOSS sample could be an outlier compared to the training set.
For $j_{\rm min} = 2$, this variability is higher than the $j_{\rm min} = 0$ case. 
As a result, we increased the number of models included in the ensemble from 10 to 50. 
We believe that this higher variability for $j_{\rm min} = 2$ is the consequence of the higher variance of large-scale information, which makes the training procedure more difficult.
We leave a precise quantification of model misspecification and strategies to resolve it in this context for future work.

Finally, we note that our results have significant discrepancies with the WST analysis of \cite{Valogiannis2022b}. We point out however that these two analyses are hardly comparable.
Besides the fact that the variant of the WST statistics used in \cite{Valogiannis2022b} significantly differ from the one we designed in this paper, the \cite{Valogiannis2022b} analysis is conducted with a Gaussian likelihood assumption following the approach of standard clustering analyses.
Furthermore, the mean function of their Gaussian likelihood is approximated by linearly extrapolating WST measurements from a fiducial cosmology using numerical derivatives, all estimated with the {\sc AbacusSummit} simulations. 
They also assume that the covariance matrix of the likelihood is independent of cosmology and use approximate {\sc Patchy} $N$-body that were calibrated using only $P_\ell$~\citep{kitaura2016}.
Our SBI analysis relaxes all of these assumptions and we rely only on full $N$-body simulations.

Overall, WST analyses present significant challenges in robustness compared to $P_\ell$ analyses.
We show that the WST is more sensitive to the model-specific imprints on galaxy clustering than the power spectrum.
In that sense, the challenges we face with robustness is in fact a demonstration of both the success of the WST and the failure of current forward models of galaxy clustering.
If forward models of galaxy clustering converged with each other at successfully describing the observed galaxy clustering, we would be able to robustly exploit the sensitivity of WST and precisely constrain the cosmological parameters.
In subsequent work, we will explore extensions of the {\sc SimBIG} forward model that add more flexibility to mitigate the impact of model misspecification.

\section{Conclusion}
\label{sec:conclusion}

We present cosmological constraints on the $\Lambda$CDM parameters $\Omega_m$, $\Omega_b$, $h$, $n_s$, and $\sigma_8$ from analyzing the BOSS CMASS SGC sample using WST statistics. Our analysis relies on a new variant of WST statistics which leverage symmetries of redshift-space data. Assuming conservative priors, we compute posterior distributions using the {\sc SimBIG} SBI inference framework. These are learned by training MAF normalizing flows on 20,000 forward modeled mock galaxy catalogs with full survey realism. The hyperparameters are optimized to maximize the best validation score, and we use an ensembling procedure of the best models to further improve this score.

We validate our posterior estimates in terms of accuracy using SBC, and in terms of robustness using the {\sc SimBIG} mock challenge, which relies on three different test sets constructed with different forward models. Our analysis passes the accuracy test but reveals robustness challenges. These take the form of a bias on $\sigma_8$ constraints when using test simulations generated from different forward models. This demonstrates that the WST statistics identify differences between the clustering of galaxies in these test sets, which are misinterpreted as variations of $\sigma_8$. We are able to improve robustness by imposing a $j_{\rm min}=2$ cut that discards small-scale clustering information beyond $k \sim 0.3~h/\text{Mpc}$.

Compared to the standard PT-based ${P_\ell(k < 0.25\,h/{\rm Mpc})}$ analysis, our WST $j_{\rm min}=2$ analysis tightens the constraints on $\Omega_m$, $\Omega_b$, $h$, $n_s$, and $\sigma_8$ by factors $1.2$, $1.5$, $0.8$, $1.3$, and $1.8$, respectively. However, despite the improved robustness of our WST $j_{\rm min} = 2$ analysis, we raise concerns on these results. We find significant variability of the posteriors across the separate models of our ensemble model. We attribute this to model misspecification. A more thorough quantification of model misspecification is still needed to further support the legitimacy of these results. We, therefore, refrain from interpreting the cosmological implication of our results. 

Our analysis underlines key challenges in leveraging new summary statistics that can extract non-Gaussian information from galaxy clustering for cosmological inference. As our WST analysis demonstrates, highly informative summary statistics are more likely to reveal model misspecification. This is a particularly important consideration for galaxy clustering where our models have not converged on small, nonlinear scales and may yet be limited in accurately describing observations. This motivates the development of even more accurate and flexible forward models as well as strategies to account for model misspecification. In subsequent work, we will analyze the WST with an extended {\sc SimBIG} forward model with more flexibility and present more reliable cosmological constraints. In accompanying papers~\citep{Hahn2024, Lemos2024}, we present parallel SimBIG analyses of galaxy clustering using the galaxy bispectrum and a field-level approach involving convolutional neural networks, respectively.

% Acknowledgements
\section*{Acknowledgments}

We thank Miles Cranmer, Mikhail Ivanov, Stéphane Mallat, Oliver Philcox, and Benjamin D. Wandelt for valuable discussions. We also acknowledge Mikhail Ivanov for providing us with posteriors used for comparison. JH has received funding from the European Union’s Horizon 2020 research and innovation program under the Marie Sk\l{}odowska-Curie grant agreement No 101025187.

\textit{Software:} {\sc Matplotlib}~\citep{Hunter2007}, {\sc nbodykit}~\citep{Hand2018}, {\sc NumPy}~\citep{Harris2020}, {\sc PyTorch}~\citep{Paszke2019}, {\sc sbi}~\citep{Tejero-Cantero2020}, {\sc SciPy}~\citep{Virtanen2020}

% Bibliography
\bibliographystyle{apsrev4-2}
\bibliography{bib/bib}

% Appendices
\begin{appendix}

\section{Isotropic spline wavelet}
\label{app:wavelet}

We define the Spherical Fourier-Bessel (SFB) transform of a square integrable scalar function $f(r, \theta, \varphi)$ as:
\begin{equation}
    \hat{f}_{lm}(k) = \sqrt{\frac{2}{\pi}}\int f(r, \theta, \varphi)j_l(kr)\conjb{Y_l^m(\theta, \varphi)}r^2\sin(\theta)\dI r\dI\theta\dI\varphi,
\end{equation}
where $Y_l^m$ functions are spherical harmonics and $j_l$ are spherical Bessel functions. The SFB transform can be inverted with the following inversion formula:
\begin{equation}
    f(r, \theta, \varphi) = \sqrt{\frac{2}{\pi}}\sum_{l=0}^{+\infty}\sum_{m=-l}^{l}\int \hat{f}_{lm}(k) k^2 j_l(kr) \dI k Y_l^m(\theta, \varphi).
\end{equation}

Following \cite{Lanusse2012}, we introduce an isotropic scaling function $\Phi$ from its SFB coefficients:
\begin{equation}
    \hat{\Phi}_{00}^{k_c}(k) = \frac{3}{2}B_3\left(\frac{2k}{k_c}\right)~\text{and}~  \hat{\Phi}_{lm}^{k_c}(k) = 0~\text{for}~l, m \neq 0,
\end{equation}
where $k_c$ is a cut-off frequency, and $B_3$ a B-spline function of order 3 defined by:
\begin{align}
    B_3(x) = \frac1{12} &\left( |x-2|^3-4|x-1|^3+6|x|^3 \right. \nonumber \\
     & \left. -4|x + 1|^3+|x + 2|^3 \right). 
\end{align}
With $j_0(x) = \sin(x)/x$ and $Y_0^0(\theta, \varphi) = (2\sqrt{\pi})^{-1}$, the previous inversion formula leads to:
\begin{equation}
    \Phi^{k_c}(r) = \begin{cases}
			\frac{768}{8\sqrt{2}\pi k_c^3r^6} &\left(4\sin(k_c r / 4) - k_c r\cos(k_c r / 4)\right) \\
                &\times \sin(k_c r/4)^3, \text{if $r \neq 0$,}\\
            \frac{k_c^3}{32\sqrt{2}\pi}, & \text{if $r = 0$.}
		 \end{cases}
\end{equation}
This function allows us to define the following isotropic spline wavelet $\psi^{\rm iso}$:
\begin{equation}
    \psi^{\rm iso}(r) = \Phi^{2k_c}(r) - \Phi^{k_c}(r).
\end{equation}
We report for reference its central frequency:
\begin{align}
    k_0 &= \int_0^{+\infty}k^3|\hat{\psi}^{\rm iso}(k)|^2\dI k / \int_0^{+\infty}k^2|\hat{\psi}^{\rm iso}(k)|^2\dI k \\
    &= \frac{10647k_c}{13394} \approx 0.8~k_c.
\end{align}

\section{Posterior estimation hyperparameter optimization}
\label{app:wandb}

\begin{table*}
    \def\arraystretch{1.5}
    \begin{tabular}{c|cccc}
        \hline
        \hline
        Hyperparameter & Minimum value & Maximum value & Distribution & Quantization step \\
        \hline
        Number of transforms & 5 & 11 & int\_uniform & N/A \\
        Number of hidden units & 256 & 1024 & q\_log\_uniform\_values & 32 \\
        Number of blocks & 2 & 4 & int\_uniform & N/A \\
        Dropout probability & 0.1 & 0.2 & q\_uniform & 0.1 \\
        Batch size & 20 & 100 & q\_uniform & 5 \\
        Learning rate & 5e-6 & 5e-5 & q\_log\_uniform\_values & 1e-06 \\
        \hline
        \hline
    \end{tabular}
    \caption{{\sc WandB} sweep configuration used for the final random search in the hyperparameter space during posterior estimation.}
    \label{table:sweep_configuration}
\end{table*}

\begin{table*}
    \def\arraystretch{1.5}
    \begin{tabular}{c|c|cccccc}
        \hline
        \hline
        Rank & Best validation & Number of & Number of & Number of & Dropout & Batch size & Learning rate \vspace{-0.25cm}\\
         & score & transforms & hidden units & blocks & probability & & \\
        \hline
        1 & 15.58 & 10 & 288 & 2 & 0.1 & 25 & 7e-06 \\
        2 & 15.56 & 6 & 864 & 2 & 0.1 & 40 & 6e-06 \\
        3 & 15.48 & 8 & 864 & 2 & 0.1 & 75 & 5e-06 \\
        4 & 15.47 & 11 & 544 & 2 & 0.1 & 40 & 5e-06 \\
        5 & 15.46 & 8 & 416 & 3 & 0.1 & 30 & 7e-06 \\
        6 & 15.44 & 6 & 704 & 3 & 0.2 & 50 & 5e-06 \\
        7 & 15.44 & 7 & 672 & 3 & 0.1 & 35 & 7e-06 \\
        8 & 15.43 & 9 & 608 & 3 & 0.1 & 50 & 7e-06 \\
        9 & 15.39 & 6 & 416 & 2 & 0.1 & 60 & 6e-06 \\
        10 & 15.38 & 6 & 416 & 2 & 0.1 & 75 & 1.2e-05 \\
        \hline
        \hline
    \end{tabular}
    \caption{Hyperparameters of the ten best architectures (for the $j_{\rm min} = 0$ analysis).}
    \label{table:best_models_hp}
\end{table*}

We give further details on the hyperparameter optimization described in Sect.~\ref{sec:posterior_estimation}.

The hyperparameters of the MAF architecture were optimized through a succession of random searches monitored by {\sc WandB}~\citep{Biewald2020}. To guarantee reproducibility of this work, we give in Table~\ref{table:sweep_configuration} the ``sweep" configuration used in our last random search, and in Table~\ref{table:best_models_hp} the hyperparameters of the ten best models (in terms of best validation score) used to define our ensemble posterior distribution. Note that these architectures were all trained using a consistent train-validation split.

We also give details on the MLP network that constitutes the conditioner of the MAF. It uses ReLU activation functions and includes batch normalization layers. These batch normalization layers significantly improved the best validation score.

\section{Additional results}
\label{app:additional_results}

\vspace{-0.5cm}
\begin{figure*}
    \includegraphics[width=0.8\textwidth]{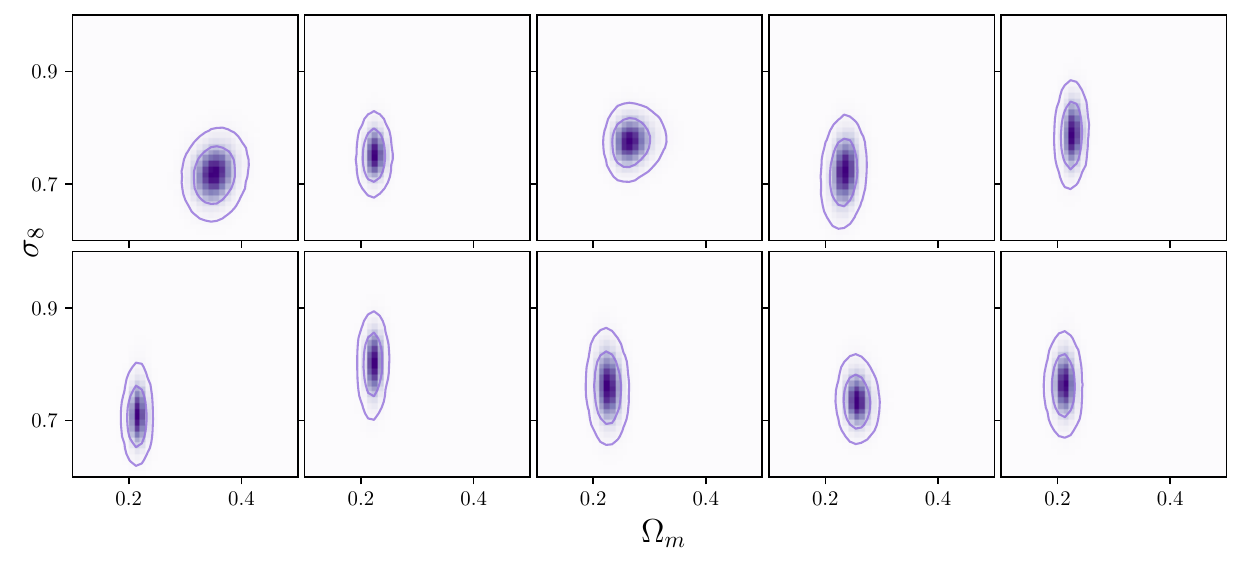}
    \caption{Observational constraints on $\sigma_8$ and $\Omega_m$ obtained across the 10 best models included in the ensemble model with $j_{\rm min} = 2$. This important variability suggests a form of model misspecification.}
    \label{fig:obs_posterior_modelmisspec_jmin_2}
\end{figure*}

\begin{figure*}
    \includegraphics[width=1\textwidth,left]{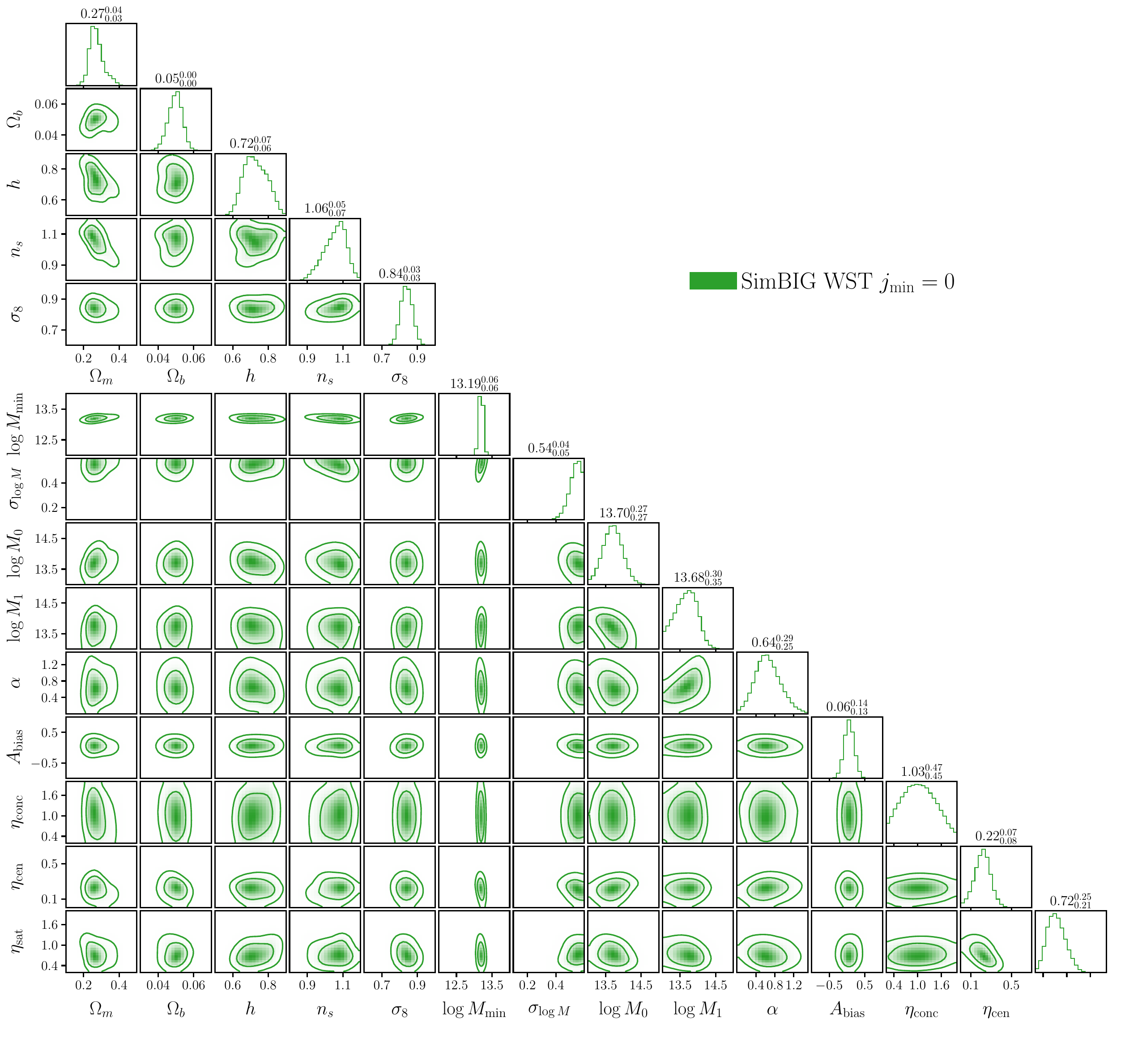}
    \caption{Same as Fig.~\ref{fig:obs_posterior}~(left) but also including predictions of the HOD parameters.}
    \label{fig:obs_full_posterior}
\end{figure*}

\begin{figure*}
    \includegraphics[width=1\textwidth,left]{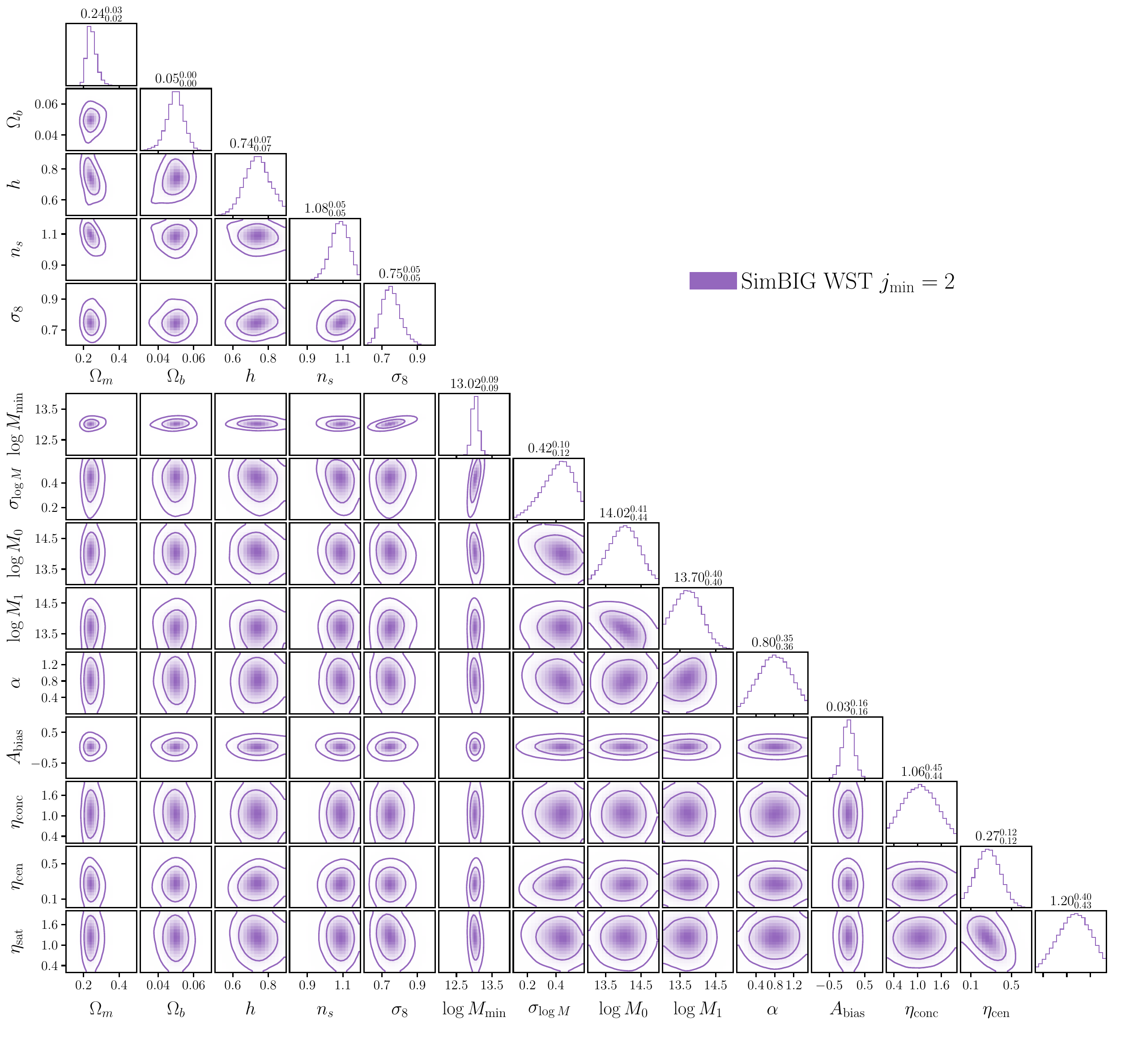}
    \caption{Same as Fig.~\ref{fig:obs_posterior}~(right) but also including predictions of the HOD parameters.}
    \label{fig:obs_full_posterior_jmin_2}
\end{figure*}

\end{appendix}

\end{document}